\documentclass[12pt,a4paper,final]{iopart}

\usepackage{iopams}  
\usepackage{graphicx}
\usepackage[breaklinks=true,colorlinks=true,linkcolor=blue,urlcolor=blue,citecolor=blue]{hyperref}

\begin{document}

\title{Spin Correlations in the Dipolar Pyrochlore Antiferromagnet Gd$_2$Sn$_2$O$_7$}

\author{Joseph A.~M. Paddison$^{1}$}
\address{$^1$School of Physics, Georgia Institute of Technology, Atlanta, GA 30332, USA}
\ead{paddison@gatech.edu}

\author{Georg Ehlers$^{2}$}
\address{$^2$Quantum Condensed Matter Division, Oak Ridge National Laboratory, Oak Ridge, TN 37831, USA}

\author{Oleg A. Petrenko$^{3}$}
\address{$^3$Department of Physics, University of Warwick, Coventry CV4 7AL, UK}

\author{Andrew R. Wildes$^{4}$}
\address{$^4$Institut Max von Laue--Paul Langevin, F-38042 Grenoble 9, France}

\author{Jason S. Gardner$^{5,6}$}
\address{$^5$Neutron Group, National Synchrotron Radiation Research Center, Hsinchu, Taiwan 30076}
\address{$^6$CCMS, National Taiwan University, Taipei, 10617, Taiwan}

\author{J. Ross Stewart$^{7}$}
\address{$^6$ISIS Facility, Rutherford Appleton Laboratory, Harwell Campus, Didcot OX11 0QX, UK}

\begin{abstract}
We investigate spin correlations in the dipolar Heisenberg antiferromagnet Gd$_2$Sn$_2$O$_7$ using polarised neutron-scattering measurements in the correlated paramagnetic regime. Using Monte Carlo methods, we show that our data are sensitive to weak further-neighbour exchange interactions of magnitude $\sim$$0.5$\% of the nearest-neighbour interaction, and are compatible with either antiferromagnetic next-nearest-neighbour interactions, or ferromagnetic third-neighbour interactions that connect spins across hexagonal loops.  Calculations of the magnetic scattering intensity reveal rods of diffuse scattering along $[111]$ reciprocal-space directions, which we explain in terms of strong antiferromagnetic correlations parallel to the set of $\langle 110\rangle$ directions that connect a given spin with its nearest neighbours. Finally, we demonstrate that the spin correlations in Gd$_2$Sn$_2$O$_7$ are highly anisotropic, and correlations parallel to third-neighbour separations are particularly sensitive to critical fluctuations associated with incipient long-range order. 
\end{abstract}

\vspace{2pc}

\section{Introduction}

\begin{figure}[htb!]
 \centering
 \includegraphics[scale=1]{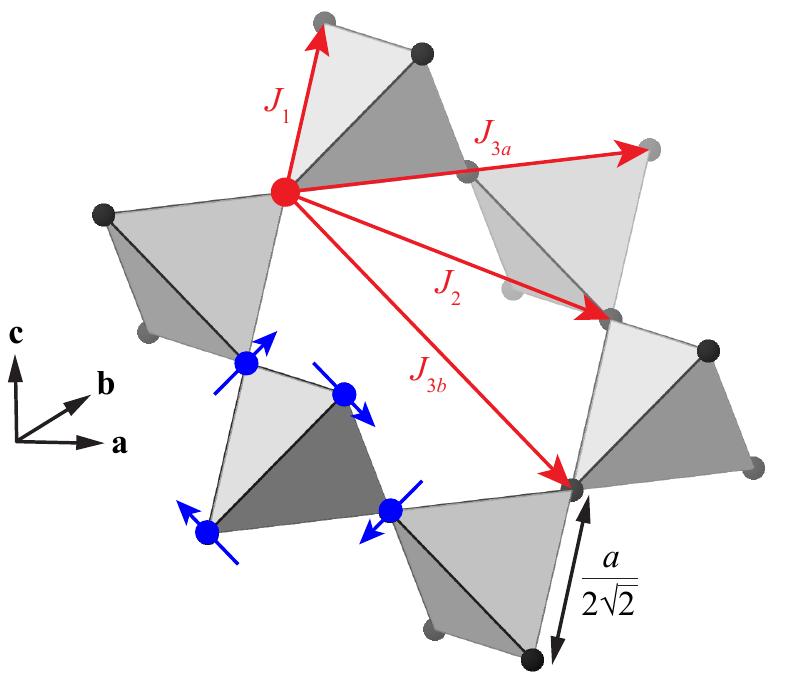}
 \caption{\label{fig:structure} Arrangement of Gd$^{3+}$ ions (black circles) in Gd$_2$Sn$_2$O$_7$. The tetrahedra of the pyrochlore lattice are shown in blue. Magnetic interaction pathways from an arbitrarily-chosen central atom (large red circle) are shown as red arrows and labelled with their exchange interaction ($J_1$, $J_2$, $J_{3a}$, and $J_{3b}$). A spin arrangement observed in Gd$_2$Sn$_2$O$_7$ below its magnetic ordering temperature (``Palmer-Chalker state") is shown by blue arrows on a single tetrahedron; the complete structure is obtained by repeating this arrangement on all tetrahedra with the same orientation. The lattice parameter $a=10.44$\,\AA.}
\end{figure}

Materials in which magnetic rare-earth ions occupy a pyrochlore lattice---a network of corner-sharing tetrahedra---have proved rewarding systems in which to study unusual types of magnetic matter. The diversity of magnetic states observed in this class of materials includes spin ices \cite{Bramwell_2001,Castelnovo_2008}, spin liquids \cite{Gardner_1999a,Fennell_2012}, fractionalised order \cite{Petit_2016}, and complex long-range order \cite{Champion_2001,Stewart_2004}. In general terms, the common origin of these states is the low connectivity of the tetrahedral building-blocks of the pyrochlore lattice [Fig.~\ref{fig:structure}]. Depending on the interplay of single-ion physics and magnetic interactions, there can be too few constraints to enforce a unique ground state, an effect called frustration \cite{Moessner_1998}. The presence of many states of similarly low energy suppresses conventional magnetic order below the Curie-Weiss temperature $\theta_\mathrm{CW}$  that defines the net strength of magnetic interactions. A canonical model that remains disordered at all temperatures contains only antiferromagnetic nearest-neighbour interactions on the pyrochlore lattice, yielding a classical spin-liquid ground state 
 \cite{Moessner_1998a,Canals_1998}. In most real materials, however, long-range magnetic order occurs at a measurable temperature $T_\mathrm{N}$, partially relieving this frustration. Even in these cases, however, materials often show a ``cooperative paramagnet" regime over a wide temperature range $T_\mathrm{N}<T\lesssim \theta_\mathrm{CW}$, where frustration has the dominant effect on the magnetic properties \cite{Moessner_1998}.

The dipolar pyrochlore antiferromagnet Gd$_2$Sn$_2$O$_7$ provides a canonical example of frustration and its relief through long-range order. The crystal structure is cubic (space group $Fd\bar{3}m$) with 16 magnetic Gd$^{3+}$ ions in the conventional unit cell \cite{Kennedy_1997}. The cubic lattice parameter $a=10.44$\,\AA~at 1.1\,K, and the lattice connectivity is shown in Fig.~\ref{fig:structure}. 
The large magnitude of the Gd$^{3+}$ spin ($S=7/2$) implies that quantum effects are minimal. It is known that there are three main terms in the spin Hamiltonian. First, an antiferromagnetic nearest-neighbour exchange interaction $J_{1}\approx -0.3$\,K is indicated by the negative value of the Curie-Weiss constant $\theta_\mathrm{CW}\approx-9.6$\,K \cite{Bondah-Jagalu_2001}. Second, the large value of $S$ implies that the long-range magnetic dipolar interaction is significant, with strength at the nearest-neighbour distance of $0.05$\,K. Third, a single-ion anisotropy term $\Delta\approx 0.14$\,K has been indicated by electron-spin resonance measurements \cite{Glazkov_2006a,Sosin_2009}, and is explained by an admixture of excited states with $L\neq0$ into the ground-state multiplet. Theoretical studies of the dipolar Heisenberg model indicate that a degeneracy of ground states exists at the mean-field level \cite{Raju_1999}; however, Palmer and Chalker showed that inclusion of higher-order terms in a free-energy expansion stablises a four-sublattice ground state with magnetic propagation vector $\mathbf{k}=(000)$ \cite{Palmer_2000}. This structure, which we will call the ``Palmer-Chalker state", is shown in Fig.~\ref{fig:structure}. In Gd$_2$Sn$_2$O$_7$, a first-order transition to an ordered state occurs at $T_\mathrm{N}=1.0$\,K ($\approx \theta_\mathrm{CW}/10$) \cite{Bonville_2003}, and neutron-diffraction measurements have shown that this is indeed the Palmer-Chalker state \cite{Wills_2006,Stewart_2008a}. Recently, debate has focussed on the possible presence of spin fluctuations at temperatures much lower than $T_\mathrm{N}$ \cite{Sosin_2009,Bertin_2002,Quilliam_2007,Bonville_2010}. Much less attention has been given to the paramagnetic phase; however, it is known to be strongly correlated, with approximately 60\% of the total magnetic entropy associated with the development of short-range correlations above $T_\mathrm{N}$ \cite{Bonville_2003}.

Here, we use neutron-scattering experiments on a polycrystalline sample of Gd$_2$Sn$_2$O$_7$ to investigate its cooperative paramagnetic phase at 1.1\,K. Our paper is structured as follows. In Section~\ref{sec:experimental}, we summarise our experimental procedures, which employed neutron polarisation analysis to isolate the magnetic scattering \cite{Scharpf_1993}.  In Section~\ref{sec:analysis}, we introduce the two methods we use to analyse our experimental data. First, we employ reverse Monte Carlo refinement, which is data-driven and ``model independent" in the sense that it does not involve a spin Hamiltonian. Second, we perform extensive Monte Carlo simulations to investigate possible models of exchange interactions in Gd$_2$Sn$_2$O$_7$ \cite{Wills_2006}. In Section~\ref{sec:results}, we explain our main results, which are as follows. We find that our diffuse-scattering data are highly sensitive to weak further-neighbour exchange interactions, and are compatible with either antiferromagnetic next-nearest-neighbour interactions $J_2$ or ferromagnetic ``cross-hexagon" interactions $J_{3b}$ that connect third neighbours across hexagonal loops [Fig.~\ref{fig:structure}]; in either case, the further-neighbour interactions are $\sim$$0.5$\% of the nearest-neighbour exchange $J_1$.  We calculate that single-crystal diffuse-scattering patterns for Gd$_2$Sn$_2$O$_7$ show prominent rods of diffuse scattering along $[111]$ reciprocal-space directions. We explain these rods in terms of strong antiferromagnetic correlations along the subset of $\langle 110\rangle$ directions that connect a given spin with its nearest neighbours. Finally, we demonstrate that the spin correlations in Gd$_2$Sn$_2$O$_7$ are highly anisotropic, and correlations parallel to one type of third-neighbour separation are particularly sensitive to the incipient long-range order. We conclude in Section~\ref{sec:conclusions} with suggestions for future work.

\section{Experimental}\label{sec:experimental}
Neutron-scattering data were collected on the same polycrystalline sample of Gd$_2$Sn$_2$O$_7$ measured previously \cite{Stewart_2008a}; the sample mass was 0.57\,g. We measured at a temperature of 1.1\,K, which is in the correlated paramagnetic regime just above $T_\mathrm{N}$. Measurements were performed using the D7 diffractometer at the Institut Laue-Langevin, Grenoble, France \cite{Stewart_2009a}. The incident neutron wavelength was 4.8\,\AA, allowing a reciprocal-space range $0.15\leq Q\leq 2.5$\,\AA$^{-1}$ to be observed. 
The technique of $xyz$ neutron polarisation analysis \cite{Scharpf_1993} was used to isolate the magnetic scattering of interest from the nuclear and spin-incoherent contributions. The data were corrected for detector and polarisation efficiency using measurements of standard samples (vanadium and amorphous silica, respectively), and were placed on an absolute intensity scale (with units of barn\,sr$^{-1}$\,Gd$^{-1}$) by normalising to the incoherent scattering from a vanadium standard.  

\section{Analysis}\label{sec:analysis}

\subsection{Theory}
An important advantage of neutron scattering is that it directly measures the spin-pair correlation function, which can allow the underlying magnetic interactions to be inferred. In our measurement, the energies of the scattered neutrons are not analysed, and the incident neutron energy (41\,K) is much larger than the sample temperature and the energy-scale of magnetic interactions ($|\theta_\mathrm{CW}|\approx 9.6$\,K \cite{Bondah-Jagalu_2001}). Under these conditions, the measurement integrates over energy transfer (``quasistatic approximation") and is sensitive to the instantaneous spin-pair correlations. The dependence of the neutron-scattering intensity on scattering vector $\mathbf{Q}$ is given by \cite{Lovesey_1987}
\begin{equation}
I(\mathbf{Q})=C[\mu f(Q)]^{2}\left[\frac{2}{3}+\frac{1}{N}\sum_{j\neq i}\mathbf{S}_{i}^{\perp}\cdot\mathbf{S}_{j}^{\perp}\exp(\mathrm{i}\mathbf{Q}\cdot\mathbf{r}_{ij})\right],
\label{eq:xtal_intensity}
\end{equation}
where $f(Q)$ is the Gd$^{3+}$ magnetic form factor \cite{Brown_2004}, $\mu =g\sqrt{S(S+1)}$ is length of the Gd$^{3+}$ magnetic moment, $C=(\gamma_{\mathrm{n}}r_{\mathrm{e}}/2)^{2}=0.07265$\,barn
is a constant, $\mathbf{r}_{ij}=\mathbf{r}_{j}-\mathbf{r}_{i}$ is the vector
connecting atoms $i$ and $j$, and $\mathbf{S}^{\perp}=\hat{\mathbf{S}}-(\hat{\mathbf{S}}\cdot\hat{\mathbf{Q}})\hat{\mathbf{Q}}$ is the component of normalised spin perpendicular to $\mathbf{Q}$. We denote vectors normalised to unit length (and operators with squared magnitude of unity) by a hat.
Since our measurement was performed on a powder sample, we spherically average (\ref{eq:xtal_intensity}) to obtain the dependence of the  intensity on the length of the scattering vector $Q=|\mathbf{Q}|$. An analytic expression for the powder-averaged intensity is given by \cite{Blech_1964,Paddison_2013}
\begin{equation}
I(Q)=C[\mu f(Q)]^{2}\left\{ \frac{2}{3}+\frac{1}{N}\sum_{j\neq i}\left[A_{ij}\frac{\sin Qr_{ij}}{Qr_{ij}}+B_{ij}\left(\frac{\sin Qr_{ij}}{(Qr_{ij})^{3}}-\frac{\cos Qr_{ij}}{(Qr_{ij})^{2}}\right)\right]\right\} ,\label{eq:powder}
\end{equation}
in which
\begin{eqnarray}
A_{ij} & = & \hat{\mathbf{S}}_{i}\cdot\hat{\mathbf{S}}_{j}-(\hat{\mathbf{S}}_{i}\cdot\hat{\mathbf{r}}_{ij})(\hat{\mathbf{S}}_{j}\cdot\hat{\mathbf{r}}_{ij}) \\
B_{ij} & = & 3(\hat{\mathbf{S}}_{i}\cdot\hat{\mathbf{r}}_{ij})(\hat{\mathbf{S}}_{j}\cdot\hat{\mathbf{r}}_{ij})-\hat{\mathbf{S}}_{i}\cdot\hat{\mathbf{S}}_{j},\label{eq:correlation_functions}
\end{eqnarray}
are spin-correlation coefficients, and $r_{ij}=|\mathbf{r}_{j}-\mathbf{r}_{i}|$. 
Importantly, the presence of the vectors $\mathbf{r}_{ij}$ in the spin-correlation coefficients shows that diffuse-scattering data are still sensitive to magnetic anisotropy after powder averaging \cite{Blech_1964}.

\subsection{Model-independent analysis}
We analyse the magnetic diffuse-scattering data for Gd$_2$Sn$_2$O$_7$
using two approaches. First, we employ reverse Monte Carlo (RMC) refinement
\cite{McGreevy_1988,Paddison_2012}, as implemented in the SPINVERT program \cite{Paddison_2013}, which uses the Metropolis algorithm
to fit spin configurations directly to the experimental diffuse-scattering
data. The ``cost function'' minimised during the refinement is given
by
\begin{equation}
\chi^{2}=W\sum_{i=1}^{N_{\mathrm{d}}}\left(\frac{I_{i}^{\mathrm{expt}}-sI_{i}^{\mathrm{calc}}}{\sigma_{i}}\right)^{2}
\end{equation}
where $I_{i}^{\mathrm{expt}}$ is the magnetic neutron-scattering intensity measured experimentally at point $i$, $I_{i}^{\mathrm{calc}}$ is the intensity calculated from (\ref{eq:powder}), $\sigma_i$ is the experimental uncertainty, $N_\mathrm{d}$ is the number of data points, $W$ is an empirical weighting
factor, and $s$ is a refined overall intensity scale factor. Refinements were initialised from random spin arrangements and were run for 500 proposed moves per spin; no further reduction in $\chi^2$ was observed after this time. A proposed move involved making the replacement 
\begin{equation}
\frac{\mathbf{S}}{|\mathbf{S}|}\rightarrow \frac{\mathbf{S}+\delta\mathbf{s}}{|\mathbf{S}+\delta\mathbf{s}|},
\label{eq:spin_move}
\end{equation}
 where $\mathbf{S}$ is a randomly-chosen spin vector, $\mathbf{s}$ is a unit vector drawn at random from the uniform spherical distribution, and $\delta=0.2$. Refinements were performed using spin configurations of size $6\times6\times6$
conventional cubic unit cells (3456 spins), and 80 separate simulations
were averaged to generate the results shown below. 
The RMC approach does not include a model of the magnetic interactions, but instead yields spin configurations
compatible with three constraints: the experimental
diffuse-scattering data, the pyrochlore lattice occupied by the Gd$^{3+}$
ions, and the fixed length of the Gd$^{3+}$ spins \cite{Paddison_2013,Paddison_2012}. Because refinements are initialised from random spin arrangements, the refined spin configurations will be as disordered as possible, provided that the constraints above are satisfied \cite{Tucker_2007}. For this reason, comparison of results from RMC refinement with predictions of interaction models can allow deficiencies in the interaction models to be identified and improved models to be built \cite{McGreevy_2001}.

\subsection{Model-dependent analysis}
We then compare the RMC results with the predictions of the spin Hamiltonian previously applied to Gd$_2$Sn$_2$O$_7$ \cite{Sosin_2009,Wills_2006}, which contains exchange interactions, a single-ion anisotropy term, and the long-range magnetic dipolar interaction:
\begin{equation}
H=-\frac{1}{2}\sum_{i,j}J_{ij}\mathbf{S}_{i}\cdot\mathbf{S}_{j}+\Delta\sum_{i}(\mathbf{S}_{i}\cdot\hat{\mathbf{z}}_{i})^2+\frac{Dr_{1}^3}{2}\sum_{i,j}\frac{\mathbf{S}_{i}\cdot\mathbf{S}_{j}-3(\mathbf{S}_{i}\cdot\hat{\mathbf{r}}_{ij})(\mathbf{S}_{j}\cdot\hat{\mathbf{r}}_{ij})}{r_{ij}^{3}}.\label{eq:hamiltonian}
\end{equation}
Here, $\mathbf{S}_{i}$ are classical vectors of magnitude $S=7/2$, $\hat{\mathbf{z}}_i$ is the local-$\langle111\rangle$ axis that connects the position of spin $i$ to the centres of the two tetrahedra which share this spin, $J_{ij}$ is the exchange interaction between spins $i$ and $j$, $\Delta$ is the single-ion anisotropy constant, and $D$ is the magnitude of the dipolar interaction at the nearest-neighbour distance $r_1$. Throughout, we fix
$D = \mu_{0}(g\mu_{\mathrm{B}})^{2}/4\pi r_{1}^{3}k_{\mathrm{B}} = 0.0496\,\mathrm{K}$, which is determined by the value of the lattice parameter ($a=10.44$\,\AA~ at 1.1\,K, from Rietveld refinement). We also fix $\Delta=0.14$\,K, as determined from electron-spin resonance measurements of the crystal-field levels  \cite{Glazkov_2006a,Sosin_2009}; this term favours spin alignment perpendicular to the local-$\langle111\rangle$ axes. 
We investigate the effects of the interactions $J_1$, $J_2$, and $J_{3b}$ shown in Fig.~\ref{fig:structure} using grid searches.
In each case, we used a Metropolis Monte Carlo algorithm to simulate (\ref{eq:hamiltonian}).
The long-range nature of the dipolar interaction was handled using
Ewald summation \cite{Wang_2001}. Simulations were run for 5000 proposed moves per spin for equilibration, starting from random spin arrangements, and snapshots were then taken every $t=500$ moves. The spin autocorrelation function was measured to check that these snapshots were essentially uncorrelated, with $\langle\mathbf{S}(0)\cdot\mathbf{S}(t)\rangle \lesssim 0.05$.  A proposed spin move was defined by (\ref{eq:spin_move}), with $\delta$ chosen so that approximately 50\% of proposed moves were accepted.
In the same way as for the RMC refinements, Monte Carlo simulations were performed using spin configurations of size $6\times6\times6$
conventional unit cells (3456 spins), and 80 separate simulations
were averaged for calculations. 

\section{Results and discussion}\label{sec:results}

\subsection{Fits}

\begin{figure}[htb!]
 \centering
 \includegraphics[scale=1]{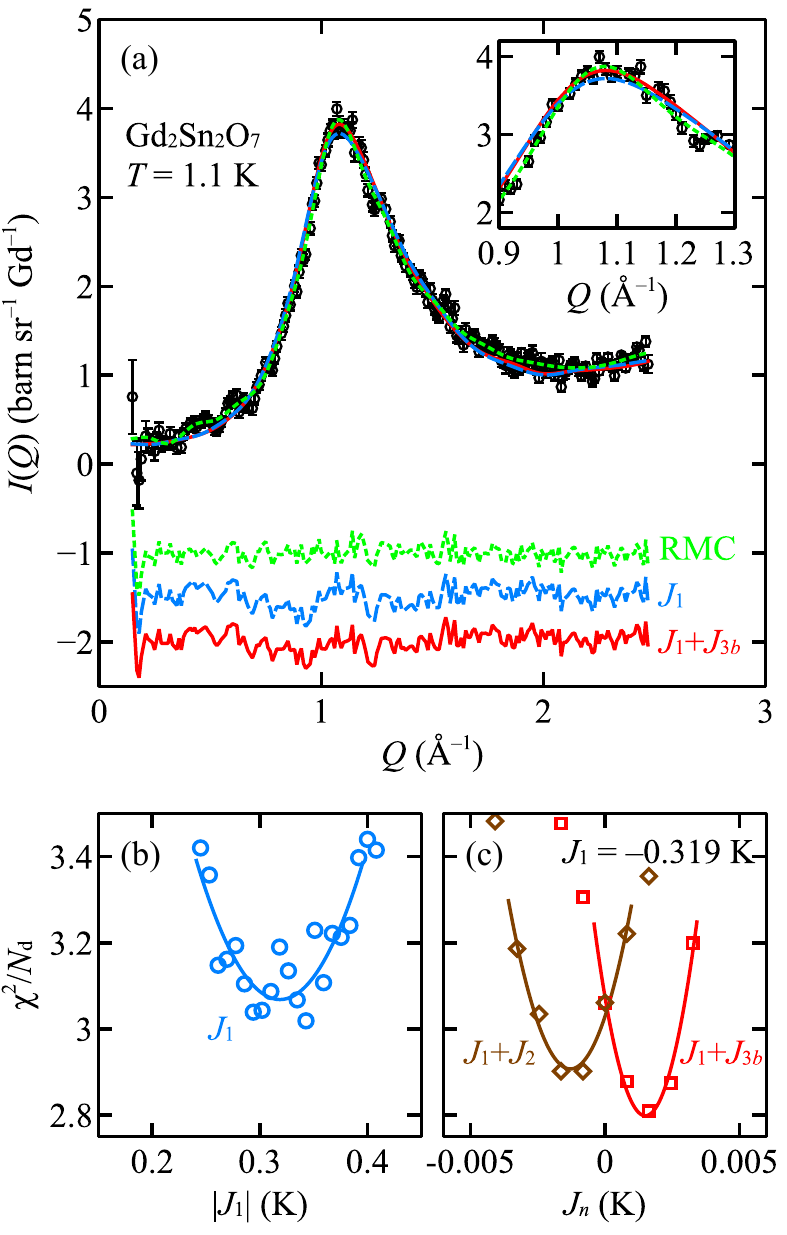}
 \caption{\label{fig:fits} (a) Experimental neutron-scattering data collected on Gd$_2$Sn$_2$O$_7$ at 1.1\,K (black circles) and fits from reverse Monte Carlo refinement (green dashed line), $J_1$-only interaction model (blue dotted line), and $J_{1}+J_{3b}$ interaction model (red solid line). The residuals (data--fit) are shown beneath the data curve (colours and line styles as above). The inset shows the diffuse peak on an expanded scale. (b) Dependence of the goodness-of-fit parameter $\chi^{2}/N_\mathrm{d}$ on the value of the nearest-neighbour interaction $J_{1}$ (blue circles). (c) Dependence of $\chi^{2}/N_\mathrm{d}$ on the value of the next-nearest-neighbour interaction $J_{2}$, keeping $J_{1}=-0.319$\,K and $J_{3b}=0$ fixed (brown diamonds), and dependence of $\chi^{2}/N_\mathrm{d}$ on the value of the third-neighbour interaction $J_{3b}$, keeping $J_{1}=-0.319$\,K and $J_{2}=0$ fixed (red squares). Solid lines are quadratic fits.}
\end{figure}

Our experimental neutron-scattering data and fits are shown in Fig.~\ref{fig:fits}(a). The main features of the data are a broad and asymmetric peak centred at $Q\approx1.1$\,\AA$^{-1}$, and a decrease in scattering intensity to approximately zero as $Q\rightarrow 0$, indicating that strong antiferromagnetic correlations persist over short distances. 
The RMC refinement yields excellent agreement with the experimental data ($\chi^{2}/N_\mathrm{d}=1.6$); such good agreement is expected because the refinement is not constrained by a specific interaction model. Turning to Monte Carlo simulations of the model Hamiltonian (\ref{eq:hamiltonian}), we first vary the value of $J_1$ to obtain the best fit, keeping all further-neighbour interactions equal to zero; we call this the ``$J_\mathrm{1}$-only model". Fig.~\ref{fig:fits}(b) shows the dependence of $\chi^{2}/N_\mathrm{d}$ on the value of $J_1$. The best fit is obtained for $J_{1} = -0.319(11)$\,K, which is similar to (but of slightly greater magnitude than) published values obtained from the magnetic susceptibility \cite{Biswas_2011} and magnetic specific heat \cite{Sosin_2009}. The quality of the fit ($\chi^{2}/N_\mathrm{d}=3.1$) is already good despite the exclusion of further-neighbour interactions, which suggests that such interactions are indeed small, as previously proposed \cite{Stewart_2008,Del-Maestro_2007}. However, the shape of the main peak is not perfectly reproduced by the $J_\mathrm{1}$-only model [inset to Fig.~\ref{fig:fits}(a)], indicating that further-neighbour interactions may still play a role. An analysis of the exchange pathways \cite{Wills_2006} suggests that the most important such interactions are likely to be $J_2$ and $J_{3b}$. We therefore investigate the effect of including either $J_2$ or $J_{3b}$ interactions in addition to $J_\mathrm{1}$, keeping $J_{1}=-0.319$\,K constant in each case.
Fig.~\ref{fig:fits}(c) shows the dependence of $\chi^{2}/N_\mathrm{d}$ on the value of $J_{2}$ or $J_{3b}$. The best fit is obtained for either antiferromagnetic values of $J_{2}$ (with $J_{3b}\equiv0$), or ferromagnetic values of $J_{3b}$ (with $J_{2}\equiv0$). We have not attempted a two-parameter fit of $J_1$ and $J_2$ or $J_{3b}$ simultaneously, due to the computational expense involved, but note that the optimal values of $J_2$ and $J_{3b}$ obtained for fixed $J_1$ are extremely small---approximately 0.5\% of $J_1$. The smaller degree of scatter evident in Fig.~\ref{fig:fits}(c) compared to Fig.~\ref{fig:fits}(b) may occur because the ordering transition is less strongly first order for non-zero $J_{2}$ or $J_{3b}$ \cite{Cepas_2005}. We interpret these results in terms of the mean-field phase diagram of the dipolar pyrochlore antiferromagnet given in \cite{Wills_2006}. For $J_{2}=J_{3b}=0$, the system lies on a boundary between competing ordering wave-vectors at the mean-field level \cite{Raju_1999}. 
For ferromagnetic $J_{2}$ or antiferromagnetic $J_{3b}$, the first ordered state has non-zero $\mathbf{k}$, whereas for antiferromagnetic $J_{2}$ or ferromagnetic $J_{3b}$---as we find for Gd$_2$Sn$_2$O$_7$---the system robustly shows $\mathbf{k}=(000)$ order at the mean-field level \cite{Wills_2006,Del-Maestro_2007}. Since Gd$_2$Sn$_2$O$_7$ actually shows $\mathbf{k}=(000)$ order, our results suggest two conclusions. First, powder-averaged magnetic diffuse-scattering data can be sensitive to very small further-neighbour exchange interactions. Second, the $\mathbf{k}=(000)$ order observed experimentally in Gd$_2$Sn$_2$O$_7$ is  stabilised by further-neighbour interactions, and not just by terms beyond mean-field level in a $J_{1}$-only model.

\subsection{Spin anisotropy}

\begin{figure}[htb!]
 \centering
 \includegraphics[scale=1]{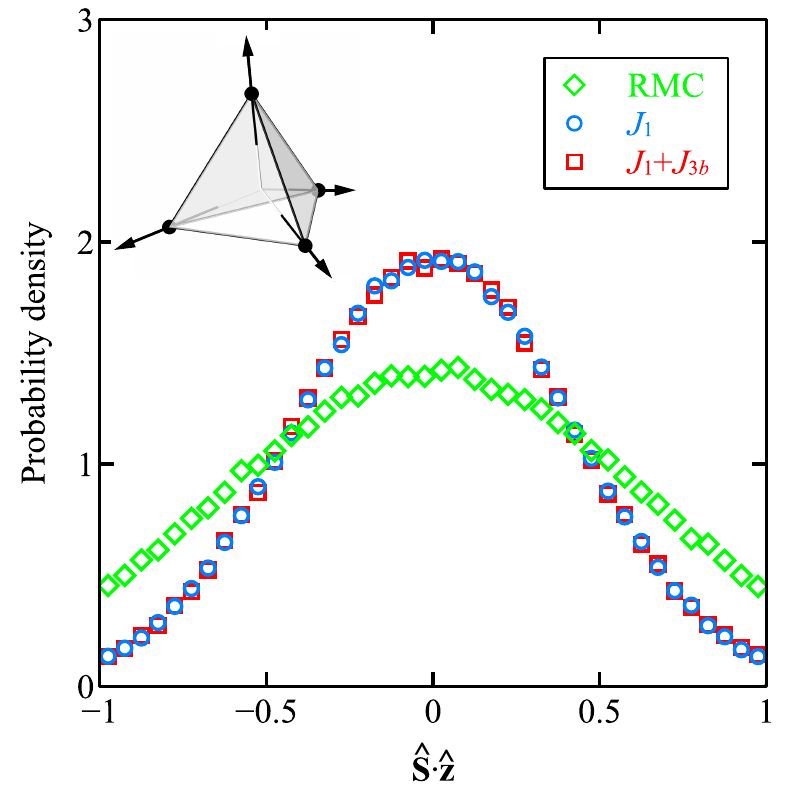}
 \caption{\label{fig:anisotropy} Probability distribution of the projection of normalised spins onto their local-$\langle 111\rangle$ axes $\hat{\mathbf{z}}$, showing results from reverse Monte Carlo refinement (green diamonds), $J_1$-only interaction model (blue circles), and $J_{1}+J_{3b}$ interaction model (red squares). The distribution function is normalised so that the area under each curve is equal to unity. The directions of the $\hat{\mathbf{z}}$ axes for each site on a tetrahedron are shown as black arrows in the inset (top left).}
\end{figure}

With both RMC refinements and interaction parameters in hand, we now investigate the correlated paramagnetic phase of Gd$_2$Sn$_2$O$_7$. In what follows, we will consider the RMC refinements, the $J_1$-only model, and the $J_{1}+J_{3b}$ model (taking $J_{3b}=0.00163$\,K). As a starting point, we consider the distribution of spin orientations with respect to the lattice. We expect that spins are oriented perpendicular to their local $\hat{\mathbf{z}}\in \frac{1}{\sqrt{3}}\langle111\rangle$ axes, because this is favoured by both the dipolar and single-ion terms. Fig.~\ref{fig:anisotropy} shows the distribution of $\hat{\mathbf{S}}\cdot{\hat{\mathbf{z}}}$ obtained from RMC refinements, and the $J_1$ and $J_{1}+J_{3b}$ models. As anticipated, all show preferential spin alignment perpendicular to $\hat{\mathbf{z}}$. Stereographic projections (not shown) revealed an isotropic distribution of spin orientations perpendicular to $\hat{\mathbf{z}}$, consistent with the absence of symmetry breaking or bond-dependent interactions.  On the one hand, the degree of anisotropy is stronger for the interaction models than for the RMC refinements, which may occur because RMC produces the most disordered (i.e., most isotropic) spin arrangements compatible with experimental data \cite{Tucker_2007}. On the other hand, the RMC results highlight that a significant degree on anisotropy is required to match the powder data shown in Fig.~\ref{fig:fits}.

\subsection{Spin correlations: reciprocal space}

\begin{figure}[htb!]
 \centering
 \includegraphics[scale=1]{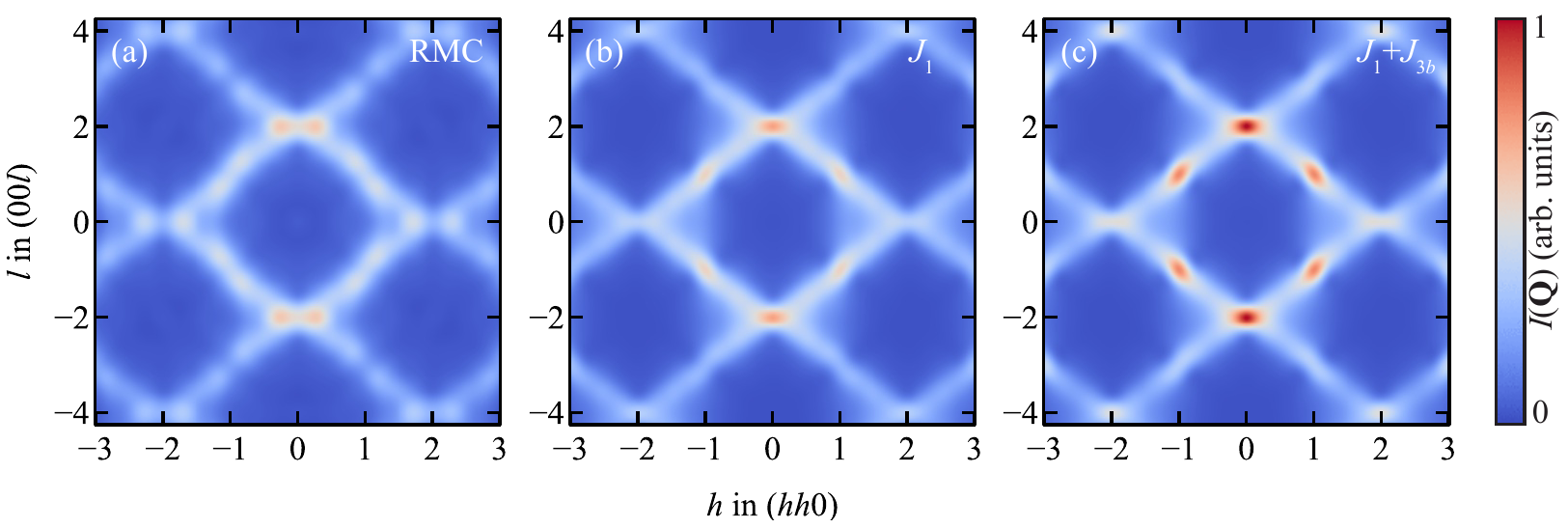}
 \caption{\label{fig:xtal} Calculated single-crystal magnetic diffuse-scattering intensity $I(\mathbf{Q})$ in the $(hhl)$ reciprocal-space plane at 1.1\,K, showing calculations from (a) reverse Monte Carlo refinement to experimental powder data; (b) $J_{1}$-only model; and (c) $J_{1}+J_{3b}$ model. All calculations are shown on the same intensity scale, and the $m\bar{3}m$ diffraction symmetry appropriate for Gd$_2$Sn$_2$O$_7$ above its magnetic ordering temperature has been applied.}
\end{figure}

We now turn to the spin-pair correlations, which we consider first in reciprocal space. While the RMC refinements are driven by fitting to powder data, calculation of the single-crystal $I(\mathbf{Q})$ is possible because a three-dimensional spin configuration is obtained; we have shown previously  \cite{Paddison_2012} that the additional constraints of fixed atomic positions and equal spin lengths means that this reconstruction is usually successful in practice. Fig.~\ref{fig:xtal}(a) shows $I(\mathbf{Q})$ from RMC refinement, and Figs.~\ref{fig:xtal}(b) and \ref{fig:xtal}(c) show $I(\mathbf{Q})$ from the $J_1$-only and $J_{1}+J_{3b}$ models, respectively. In all cases, the scattering in the $(hhl)$ plane was calculated from the relevant spin configurations by applying (\ref{eq:xtal_intensity}). The dominant features---observed in both RMC refinements and model calculations---are rods of diffuse-scattering intensity along $[111]$ reciprocal-space directions. We confirmed that this description in terms of rods is correct by calculating $I(\mathbf{Q})$ in all three dimensions of reciprocal space. The similarity of the results from RMC refinements and model simulations [Fig.~\ref{fig:xtal}] indicates that it is very likely that rods of diffuse scattering would be observed if experiments were performed on single-crystal samples of Gd$_2$Sn$_2$O$_7$.
In addition to the rods themselves, diffuse peaks occur at positions where several rods intersect, which are also the positions of magnetic Bragg peaks in the Palmer-Chalker state. The strongest such peaks occur at $\{002\}$ positions, where four rods intersect (two of which lie within the $(hhl)$ plane shown in Fig.~\ref{fig:xtal}). Weaker peaks are observed at $\{111\}$ positions, where two rods intersect (one of which lies within the $(hhl)$ plane). The $\{111\}$ and $\{002\}$ peaks are strongest for the $J_{1}+J_{3b}$ calculation, which may indicate the development of critical fluctuations associated with antiferromagnetic $\mathbf{k}=(000)$ order, as we discuss in more detail below. Interestingly, very similar $[111]$ rods of diffuse scattering were observed in the ferromagnetic ``quantum spin ice" candidate Yb$_2$Ti$_2$O$_7$, with the important difference that in that material the rods intersect at the $(000)$ and $\{222\}$ positions \cite{Ross_2009,Thompson_2011,Chang_2012}---i.e., those associated with ferromagnetic $\mathbf{k}=(000)$ order.

\subsection{Spin correlations: real space}

\begin{figure}[htb!]
 \centering
 \includegraphics[scale=1]{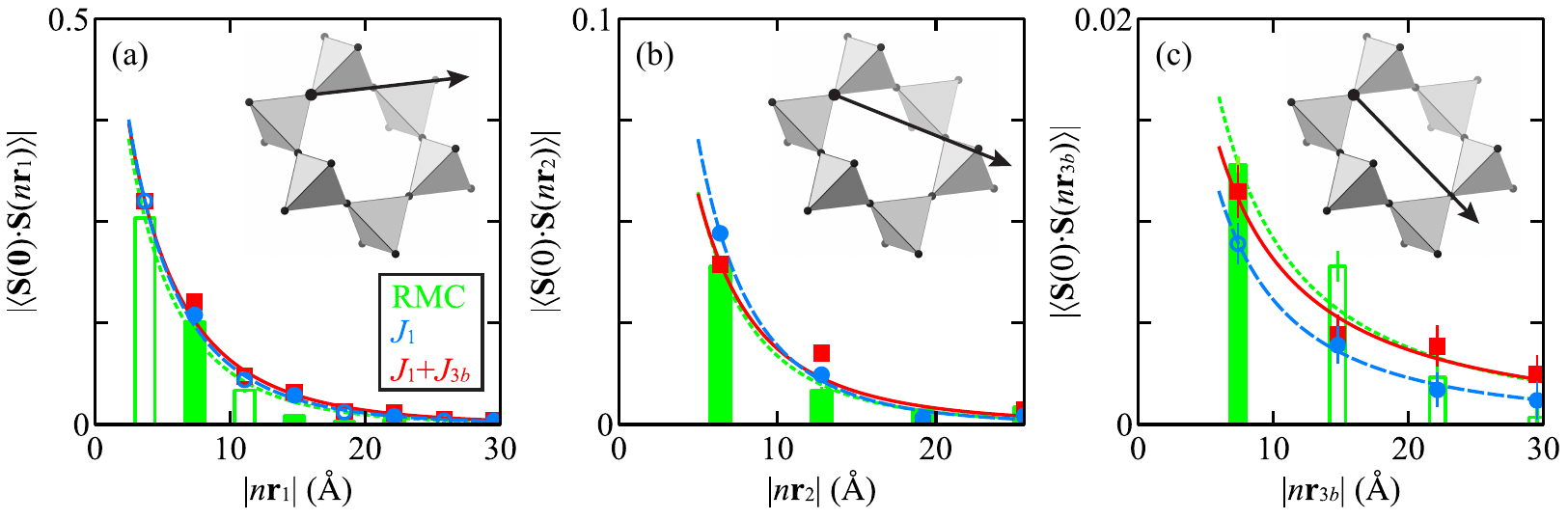}
 \caption{\label{fig:correlations} Spin correlation functions along high-symmetry directions in real space, showing (a) correlations parallel to nearest-neighbour vectors $\mathbf{r}_1$; (b) correlations parallel to next-nearest-neighbour vectors $\mathbf{r}_2$; and (c) correlations parallel to cross-hexagon vectors $\mathbf{r}_{3b}$. In (a), (b), and (c), the direction along which spin correlations are plotted is indicated by the black arrow in the diagram (top right); note the different vertical scale in each panel. Results from reverse Monte Carlo refinements to experimental data are shown as green bars, calculations from the  $J_{1}$-only model are shown as blue circles, and calculations from the  $J_{1}+J_{3b}$ model are shown as red squares. Ferromagnetic (positive) correlations are shown as solid bars or symbols, and antiferromagnetic (negative) correlations as hollow bars or symbols. Lines show stretched-exponential fits (colours as above).}
\end{figure}

To explain the diffuse-scattering features shown in Fig.~\ref{fig:xtal}, we consider the spin correlations in real space. In general, one-dimensional features (rods) in reciprocal space are generated by two-dimensional (planar) correlations in real space, where the real-space plane is perpendicular to the reciprocal-space rod. Hence, we look for correlations in $\{111\}$ planes in real space. 
Each of the four $\{111\}$ planes contains 50\% of the nearest-neighbour vectors $\mathbf{r}_{1}$, 25\% of the next-nearest-neighbour vectors $\mathbf{r}_{2}$, and 50\% of the cross-hexagon vectors $\mathbf{r}_{3b}$. We therefore expect that the diffuse rods may be associated with strong correlations along $\mathbf{r}_{1}$ and/or $\mathbf{r}_{3b}$ directions, because these are the most strongly represented directions within $\{111\}$ planes. To test this hypothesis, we define the spin correlation function
\begin{equation}
\langle \mathbf{S}(\mathbf{0})\cdot\mathbf{S}(\mathbf{r})\rangle =\frac{1}{N}\sum_{i=1}^{N}\sum_{j=1}^{Z_{\mathbf{r}}}\frac{\hat{\mathbf{S}}_{i}\cdot\hat{\mathbf{S}}_{j}}{Z_{\mathbf{r}}},
\end{equation}
which is the scalar product of a normalised spin with its neighbour at vector separation $\mathbf{r}$, averaged over the $Z_\mathbf{r}$ symmetry-equivalent neighbours and $N$ spins as centres. 
Fig.~\ref{fig:correlations} shows $\langle \mathbf{S}(\mathbf{0})\cdot\mathbf{S}(\mathbf{r})\rangle$ parallel to nearest-neighbour, next-nearest neighbour, and cross-hexagon directions (i.e., $\mathbf{r}=n\mathbf{r}_{1}$, $n\mathbf{r}_{2}$, and $n\mathbf{r}_{3b}$, respectively, where $n$ is an integer). The sign of $\langle \mathbf{S}(\mathbf{0})\cdot\mathbf{S}(n\mathbf{r}_{1})\rangle$ alternates according to $(-1)^n$---a likely consequence of the dominant antiferromagnetic nearest-neighbour exchange interactions---whereas the sign of $\langle \mathbf{S}(\mathbf{0})\cdot\mathbf{S}(n\mathbf{r}_{2})\rangle$ is positive for all $n$. Over comparable distances, the correlations parallel to $\mathbf{r}_1$ are of much greater magnitude than those parallel to $\mathbf{r}_2$ and $\mathbf{r}_{3b}$.  
These trends are consistent across the RMC, $J_1$-only, and $J_{1}+J_{3b}$ models;
taking the $J_1$-only model as an example, the ratios $\left|\frac{\langle \mathbf{S}(\mathbf{0})\cdot\mathbf{S}(\mathbf{r}_{1})\rangle}{\langle \mathbf{S}(\mathbf{0})\cdot\mathbf{S}(\mathbf{r}_{2})\rangle}\right|=5.85(6)$, and $\left|\frac{\langle \mathbf{S}(\mathbf{0})\cdot\mathbf{S}(2\mathbf{r}_{1})\rangle}{\langle \mathbf{S}(\mathbf{0})\cdot\mathbf{S}(\mathbf{r}_{2})\rangle}\right|=2.86(4)$. It is interesting to compare our results for Gd$_2$Sn$_2$O$_7$ with a model containing nearest-neighbour Heisenberg exchange interactions only (i.e., $D=0$ and $\Delta=0$ in (\ref{eq:hamiltonian})). This model is a paradigm of frustrated magnetism: its scattering pattern shows ``pinch point" features and does \emph{not} show rods of diffuse scattering  \cite{Moessner_1998,Moessner_1998a,Isakov_2004,Henley_2005}. Our Monte Carlo simulations of this model at $T\ll J$ reveal that it shows much smaller ratios of 2.733(9) and 0.732(7), respectively.
We therefore suggest that rods of diffuse scattering in Gd$_2$Sn$_2$O$_7$ are a consequence of strong antiferromagnetic correlations along $\mathbf{r}_1$ directions. 

We now ask whether the incipient $\mathbf{k}=(000)$ ordering is evident in real space. As noted above, the correlations along $\mathbf{r}_1$ and $\mathbf{r}_2$ directions are similar for all the models we consider. Correlations along the $\mathbf{r}_{3b}$ directions are always very small, probably because they are particularly strongly frustrated: the Palmer-Chalker structure shows ferromagnetic spin alignment at $\mathbf{r}_{3b}$ [Fig.~\ref{fig:structure}], whereas antiferromagnetic $J_1$ favours antiferromagnetic alignment because $\mathbf{r}_{3b}=3\mathbf{r}_1$. Despite their small magnitudes, the correlations along $\mathbf{r}_{3b}$ differ strongly between models. The RMC model shows ferromagnetic correlation at $\mathbf{r}_{3b}$ followed by antiferromagnetic correlations at $n\mathbf{r}_{3b}$ with $n>1$; the $J_1$-only model shows the opposite trend; and the $J_1+J_{3b}$ model shows ferromagnetic correlations at all $n\mathbf{r}_{3b}$. Hence, $ \langle \mathbf{S}(\mathbf{0})\cdot\mathbf{S}(n\mathbf{r}_{3b})\rangle$ for the $J_1+J_{3b}$ model resembles the Palmer-Chalker state, and also resembles the RMC result more closely than the $J_1$-only model. These results suggest that spin correlations along $\mathbf{r}_{3b}$ directions can be particularly sensitive to the structure-directing effects of further-neighbour interactions. To assess this effect in more detail, we fitted $\left| \langle \mathbf{S}(\mathbf{0})\cdot\mathbf{S}(\mathbf{r})\rangle \right|$ along different directions. The correlation function does not follow a simple exponential decay, but can be reasonably well described by a stretched-exponential form,
$\left| \langle \mathbf{S}(\mathbf{0})\cdot\mathbf{S}(\mathbf{r})\rangle \right|=\exp \left[-\left(\frac{|\mathbf{r}|}{\xi}\right)^{\beta}\right]$, with different values of the correlation length $\xi$ and stretching exponent $\beta$ along each direction. As anticipated, the correlation lengths decrease in the order $\xi(\mathbf{r}_{1})>\xi(\mathbf{r}_{2})\gg \xi(\mathbf{r}_{3b})$; e.g., for the $J_1$-only model, we obtain values of $2.582(4)$\,\AA, $0.90(5)$\,\AA, and $0.02(2)$\,\AA, respectively. The stretching exponents also decrease in the order $\beta(\mathbf{r}_{1})>\beta(\mathbf{r}_{2})\gg \beta(\mathbf{r}_{3b})$; e.g., for the $J_1$-only model, we obtain exponents of $0.710(2)$, $0.57(2)$, and $0.26(5)$, respectively. Physically, this means that correlations along $\mathbf{r}_{1}$ directions are most similar to the exponential decay expected in a conventional paramagnet, whereas correlations along $\mathbf{r}_{3b}$ directions rapidly decay to a value close to zero but have a ``long tail" at large distances. It is natural to identify the correlations along $\mathbf{r}_{3b}$ directions with critical fluctuations, because they have a long-range component and are sensitive to the incipient ordered state. Hence, our results suggest that correlations in Gd$_2$Sn$_2$O$_7$ are highly anisotropic in real space, with critical fluctuations occurring selectively along $\mathbf{r}_{3b}$ directions. These results are very different to Yb$_2$Ti$_2$O$_7$, which shows nearly isotropic behaviour of the spin-correlation function despite also showing rods of diffuse scattering \cite{Thompson_2011}, a difference that may perhaps be related to the much weaker dipolar interaction in Yb$_2$Ti$_2$O$_7$ \cite{Ross_2011}.  

\section{Conclusions}\label{sec:conclusions}

Our results suggest several avenues for future work. First, single-crystal neutron scattering experiments would allow a direct experimental test of our prediction of rod-like diffuse-scattering features. Such experiments are challenging, however, because of the difficulty of preparing large single crystals of Gd$_2$Sn$_2$O$_7$ and the large neutron-absorption cross-section of natural Gd. Our observation that the powder diffuse-scattering profile is sensitive to the weak further-neighbour interactions that select a particular ordered state may be relevant to the related materials Gd$_2$Pb$_2$O$_7$ \cite{Hallas_2015}, Gd$_2$Pt$_2$O$_7$ \cite{Hallas_2016}, and Gd$_2$Ti$_2$O$_7$, in which the nature of the magnetic order is not yet conclusively established \cite{Stewart_2004,Glazkov_2006,Glazkov_2007}. Careful measurement of the powder $I(Q)$ could allow accurate determination of the further-neighbour exchange interactions, allowing the nature of long-range order to be predicted in these systems. This may be more straightforward for Gd$_2$Ti$_2$O$_7$, because its smaller lattice constant (10.17\,\AA~\cite{Stewart_2004} \emph{vs.} 10.44\,\AA~in Gd$_2$Sn$_2$O$_7$) indicates that further-neighbour exchange interactions should play a more important role. Finally, a fuller understanding of the weak further-neighbour interactions Gd$_2$Sn$_2$O$_7$ may prove important to model the complex sequence of magnetic phase transitions it exhibits as a function of applied magnetic field \cite{Freitas_2011}.

\section*{Acknowledgements}
We are grateful to Andrew Goodwin and Martin Mourigal for valuable discussions. The work at the Georgia Institute of Technology (J.A.M.P.) was supported by the College of Sciences and the Executive Vice-President for Research. G.E. acknowledges funding by the Scientific User Facilities Division, Office of Basic Energy Sciences, U.S. Department of Energy. We gratefully acknowledge the ILL cryogenics team for technical support.

\section*{References}
\bibliographystyle{Science_allauthors.bst}
\bibliography{jamp_full_refs}

\end{document}